\documentclass[aps,pra,twocolumn,superscriptaddress]{revtex4-1}

\usepackage{graphicx}
\usepackage{amsmath}
\usepackage{amsfonts}
\usepackage{relsize}
\usepackage{amssymb}

\newcommand{\ket}[1]{|#1 \rangle}
\newcommand{\bra}[1]{\langle #1|}

\begin{document}
	
\title{Approximate solutions to Mathieu's equation}	
\author{Samuel A. Wilkinson}
\affiliation{Chemical and Quantum Physics, School of Science, RMIT University, Melbourne, Victoria 3001, Australia}
\author{Nicolas Vogt}
\affiliation{Chemical and Quantum Physics, School of Science, RMIT University, Melbourne, Victoria 3001, Australia}
\author{Dmitry S. Golubev}
\affiliation{Low Temperature Laboratory, Department of Applied Physics, Aalto University School of Science, P.O. Box 13500, FI-00076 Aalto, Finland}
\author{Jared H. Cole}
\affiliation{Chemical and Quantum Physics, School of Science, RMIT University, Melbourne, Victoria 3001, Australia}
	
\date{\today}
	
\begin{abstract}
	Mathieu's equation has many applications throughout theoretical physics. It is especially important to the theory of Josephson junctions, where it is equivalent to Schr\"{o}dinger's equation. Mathieu's equation can be easily solved numerically, however there exists no closed-form analytic solution. Here we collect various approximations which appear throughout the physics and mathematics literature and examine their accuracy and regimes of applicability. Particular attention is paid to quantities relevant to the physics of Josephson junctions, but the arguments and notation are kept general so as to be of use to the broader physics community.
\end{abstract}
	
\maketitle

\section{Introduction}	
Mathieu's equation,
\begin{equation} \label{eq:Mathieu}
\frac{d^2 \psi}{dz^2} + (a - 2\eta\cos(2z))\psi = 0.
\end{equation}
has appeared in theoretical physics in many different contexts. 
Mathieu originally formulated the equation to describe the vibration modes of an elliptical membrane \cite{Mathieu1868}, but the equation has since been applied to the theory of quadrupole ion traps \cite{March1997,Konenkov2002,Baranov2003}, ultracold atoms \cite{Rey2005} and quantum rotor models \cite{Ayub2009,Condon1991}. 
This equation has also found attention as a simplified model of a particle moving in a periodic potential \cite{Slater1952}. 

Although Mathieu's equation is easy to solve numerically, and although exact results are achievable in certain limits, a general analytic solution of Mathieu's equation has not yet been achieved.
Instead, there exists throughout the literature, both on physics and mathematics, a myriad of approximations and numerical methods which may be used to extract quantities of interest.
It is the goal of this paper to collect these approximations together in one place for easy reference, to review them explicitly and explore their regimes of validity.
The focus is to illustrate and compare the results found in the vast body of literature on this topic.

This manuscript will focus primarily on applications of Mathieu's equation to the physics of Josephson junctions \cite{Likharev1985,Koch2007,Harbaugh2000,Hermon1996}, however we will keep the notation general as the results presented herein may be of use across diverse fields.
Josephson junctions are elements in superconducting circuits, which are of great interest due to potential applications in quantum technology \cite{Likharev1985,Makhlin2001,Martinis2009}.

A single Josephson junction is governed by the Hamiltonian
\begin{equation}
H = -4E_C\frac{\partial^2}{\partial\phi^2} - E_J\cos(\phi)
\end{equation}
where $E_C = e^2/2C$ is the charging energy, $C$ is the junction capacitance, $E_J$ the Josephson energy and $\phi$ is the phase difference of the superconducting condensate across the junction. 
With this Hamiltonian, the time-independent Sch\"{o}dinger equation becomes
\begin{equation}
\left[-4E_C\frac{\partial^2}{\partial\phi^2} - E_J\cos(\phi)\right]\psi = E\psi.
\end{equation}
This reduces to Mathieu's equation upon making the substitutions $\phi/2 \rightarrow z$, $E/E_C \rightarrow a$, $E_J/2E_C \rightarrow \eta$. 
To maintain generality, we will retain the notation of Mathieu's equations, but we will bear these substitutions in mind and make frequent reference to results obtained in the theory of Josephson junctions. 

\begin{figure}
	\centering
	\includegraphics[width=1\linewidth]{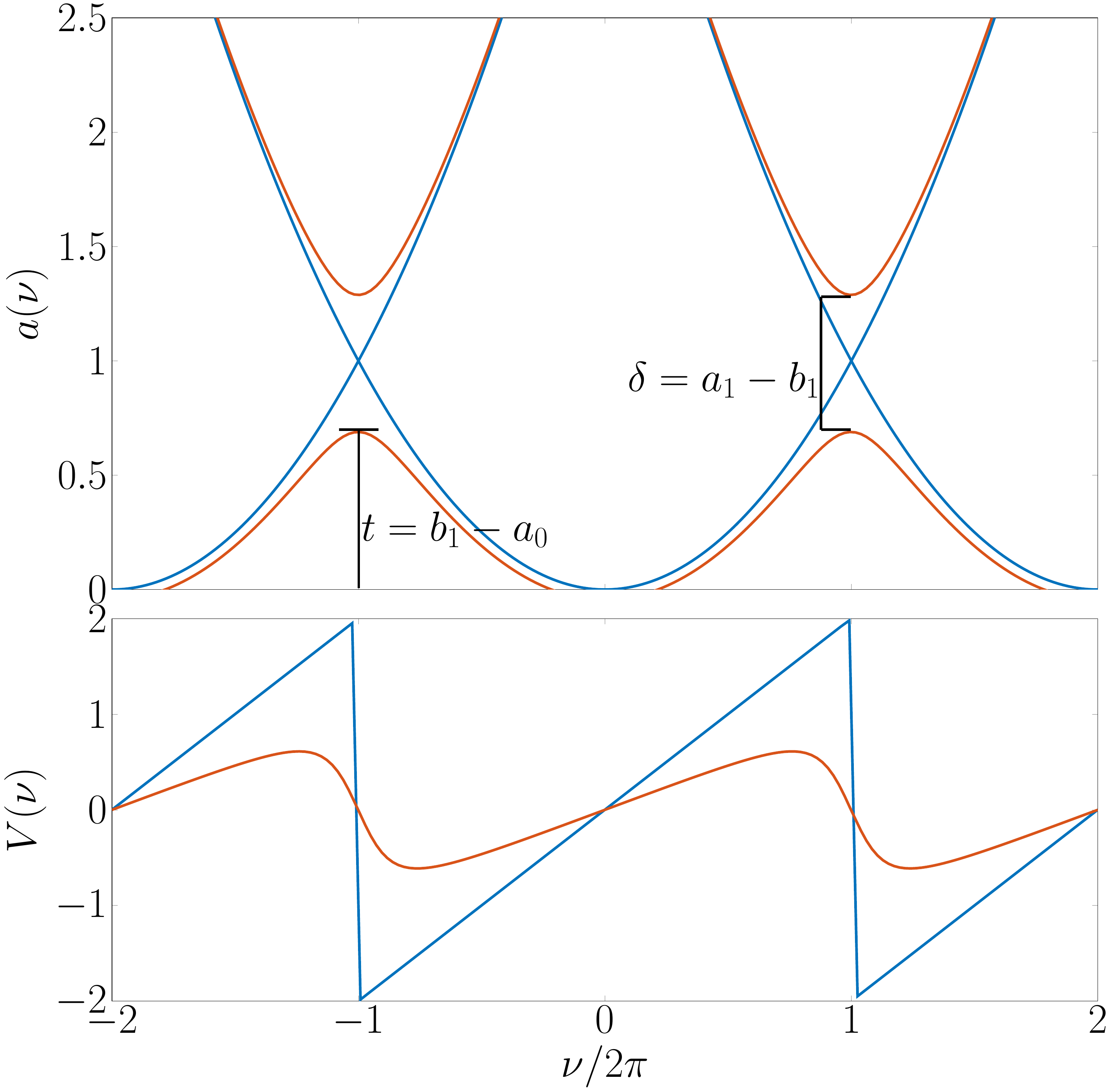}
	\caption{\label{fig:BandsLowEta} The characteristic value $a(\nu)$ and its derivative with respect to the Floquet exponent $\nu$ for Mathieu's equation with $\eta = 0$ (blue) and $\eta = 0.3$ (red).}
\end{figure}

The focus will be on quantities corresponding to physical observables in Josephson junctions.
We will therefore not be concerned with the details of the Mathieu functions themselves (physically, the wavefunctions of the Josephson junction array), but primarily on the characteristic value $a$, the floquet exponent $\nu$, and related quantities depicted in Fig.~\ref{fig:BandsLowEta}.

Each of these quantities will be discussed in detail below, but each can be understood loosely as follows: $t = b_1 - a_0$ is the difference between the lowest characteristic value of an odd-parity Mathieu function and the lowest characteristic value of an even-parity Mathieu function. Physically it corresponds to the bandwidth of the lowest energy band of a Josephson junction.

For characteristic values betweem $a_1$ and $b_1$, stable Mathieu functions do not exist.
$\delta = a_1 - b_1$ represents a gap in characteristic values of stable Mathieu functions.
Physically, $\delta$ corresponds to the band gap in the energy spectrum of the Josephson junction.

$V(\nu) = \textrm{d}a/\textrm{d}\nu$ is a quantity little discussed in the mathematics literature, but in the physics of Josephson junctions it is known as the effective voltage \cite{Likharev1985}.

In experiments on Josephson junctions the quantity $\eta$ is often a controlled parameter.
In fact, if one adopts a SQUID geometry, $E_J$, and by extension $\eta$, can be tuned in real time by adjusting the applied magnetic flux \cite{Zagoskin2011}.
We are therefore primarily interested with how these various parameters vary with $\eta$.
In Fig.~\ref{fig:BandsLowEta} we have ploted $a(\nu)$ and $V(\nu)$ for $\eta = 0$ and $\eta = 0.2$.

The limits of both strong coupling ($\eta \gg 1$) and weak coupling ($\eta \ll 1$) are relatively straightforward. 
In both cases the characteristic values can be expressed as asymptotic expansions in powers of $\eta$ or $1/\eta$ respectively.
Below we will explore both of these extreme limits of the model, and investigate the region $\eta \sim 1$ where the approximations are expected to break down.
We will also examine properties of Mathieu's equation which may be deduced from periodicity arguments, as these are expected to be valid for any value of $\eta$.

\section{Small $\eta$}
In the limit that $\eta\rightarrow 0$, Mathieu's equation becomes
\begin{equation}
\frac{\textrm{d}^2 \psi}{\textrm{d}\phi^2} + a\psi = 0.
\end{equation}
This differs from Schr\"{o}dinger's equation for a particle moving in free space only in that the co-ordinate $\phi$ has the topology of a circle. 
In this limit, the eigenvalues are continuous and do not form separate energy bands or levels.
The Mathieu functions themselves are simply $\pm\cos(\sqrt{a_n} z)$, $\pm\sin(\sqrt{b_{n+1}}z)$ (as can be trivially verified).
By convention we take the sign to be positive.
The characteristic value of the $\sin$ solution is denoted $b_{n+1}$ rather than $a_n$ by convention and for later convenience, but it should be interpreted the same way (physically, as an energy eigenvalue).

For finite $\eta$ corrections must be added to the simple $\cos$ and $\sin$ solutions, however the solutions retain their periodicity and parity.
The finite $\eta$ generalisations are referred to as cosine-elliptic or sine-elliptic functions respectively, and are denoted $\textrm{ce}_{n}(z,\eta)$ and $\textrm{se}_{n+1}(z,\eta)$.
These can generally not be expressed in closed form.
However, we can obtain many physically relevant quantities without direct reference to these functions.

At $\eta=0$, stable solutions exist for any value of $a_n$ (or $b_m$).
However, at finite $\eta$ band gaps appear, and solutions are only stable when the characteristic value $a$ is $a_n \leq a \leq b_{n+1}$, where $n$ is an integer and where we have used $a$ without a subscript to denote an arbitrary characteristic number which will generally be of fractional order.

Physically, this stability/instability of solutions manifests itself in the form of energy bands, so that the stability diagram of Mathieu's equation gives us the band structure of a Josephson junction.
At a given value of $\eta$, the characteristic energy is a periodic function of the characteristic exponent $\nu$ (to be introduced below).
Many quantities of physical interest can be expressed in terms of the lowest and highest energies in a band, $a_n$ and $b_{n+1}$ respectively. 
For example, the ground state bandwidth is just $b_1 - a_0$, and the gap between the ground and first excited state is $a_1 - b_1$.

At small $\eta$, the characteristic values can be expanded in powers of $\eta$ \cite{McLachlan1947}, giving
\begin{equation}
\begin{split} \label{eq:McLachlan}
a_0 =& -\frac{1}{2}\eta^2 + \frac{7}{128}\eta^4 - \frac{29}{2304}\eta^6 + \frac{68687}{18874368}\eta^8 + \mathcal{O}(\eta^{10})\\
b_1 =& 1 - \eta - \frac{1}{8}\eta^2 + \frac{1}{64}\eta^3 - \frac{1}{1536}\eta^4 - \frac{11}{36864}\eta^5 + \frac{49}{589824}\eta^6\\ & - \frac{55}{9437184}\eta^7 - \frac{265}{113246208}\eta^8 + \mathcal{O}(\eta^9).
\end{split}
\end{equation}
The expression for $a_1$ is identical to $b_1$, but with $\eta \rightarrow -\eta$.
Similar expansions for higher order characteristic values can be found in section 2.151 of ref.~\cite{McLachlan1947}.

\section{Large $\eta$}
When $\eta \gg 1$, $z$ remains close to the minima of $\cos2z$, so that when expanded as a Taylor series only the second order term is relevant. This reduces the Mathieu equation to the form of Schr\"{o}dinger's equation for a harmonic oscillator, so that the Mathieu functions may be approximated by the wavefunctions of a harmonic oscillator
\begin{equation}
\psi_n^\textrm{HO}(z) = c_n H_n\left((2\eta)^{1/4}z\right)e^{-\frac{1}{2}\sqrt{2\eta}z^2}
\end{equation}
with energy levels
\begin{equation}
a_n = 4\sqrt{\eta}(n + \frac{1}{2}) -2\eta
\end{equation}
where $H_n(x)$ are Hermite polynomials familiar from the theory of the quantum harmonic oscillator, $c_n$  is a normalization constant and the constant shift $2\eta$ comes from the expansion of the cosine. Introducing $x = (2\eta)^{1/4}z$ this simplifies to
\begin{equation} \label{eq:HO}
\psi_n^\textrm{HO}(x) = c_n H_n\left(x\right)e^{-\frac{1}{2}x^2}.
\end{equation}

In this limit, the Mathieu equation can be interpreted as the Hamiltonian for a tight-binding model \cite{Grosso2014}.
Following the standard textbook analysis of the tight-binding model, we can calculate the bandwidth of the characteristic values of the Mathieu equation via
\begin{equation}
b_1 - a_0 = -\int \textrm{d}z \psi(z)\psi(z - \pi)V(z)
\end{equation}
where $V(z) = -2\eta(1+\cos(z)) \approx \eta z^2 + \textrm{const.}$ and we have shifted our integration variable $2z \rightarrow z$.
A detailed calculation of this integral, along with a discussion on the appropriate approximate wave-functions is given in \cite{Catelani2011}.
The final result is
\begin{equation} \label{eq:TBbandwidth}
b_1 - a_0 = 16\sqrt{\frac{2}{\pi}}\eta^{3/4}e^{-4\sqrt{\eta}},
\end{equation}
or, expressed in the notation relevant to Josephson junctions \cite{Likharev1985,Averin1991a},
\begin{equation}
t_0 = 32 \left(\frac{E_JE_C}{\pi}\right)^{1/2} \left(\frac{E_J}{2E_C}\right)^{1/4} \exp\left[-\left(8\frac{E_J}{E_C}\right)^{1/2}\right].
\end{equation}
The exponential decay of the bandwidth with $\sqrt{\eta}$ justifies the approximation of the bands as infinitely thin at large $\eta$ in the asymptotic expansions of $a_n$.

Higher order corrections are given in section 3.43 of \cite{McLachlan1947}, however the corrections are only polynomial in $\eta$ so for large $\eta$ the exponential decay is the dominant feature.
	
As the bandwidth shrinks exponentially, the characteristic values $a_n$ and $b_{n+1}$ become approximately equal at large $\eta$.
We can therefore neglect the difference between the two and treat the bands as being infinitely thin, corresponding to a single energy which for convenience we will label $a_n$.
Asymptotic expansions for this value exist \cite{Frenkel2001}, and we find that $a_n$ to order $\eta^{-1}$ is
\begin{equation}
\begin{split} \label{eq:F+P}
a_n = &-2\eta + (2+4n)\sqrt{\eta} - \frac{1}{4} - \frac{1}{2}n - \frac{1}{2}n^2 \\
&+ \left(-\frac{1}{32} - \frac{3}{32}n - \frac{3}{32}n^2 - \frac{1}{16}n^3\right)\frac{1}{\sqrt{\eta}} \\
&+ \left(-\frac{11}{256} -\frac{3}{256}n - \frac{1}{16}n^2 - \frac{5}{256}n^4\right)\frac{1}{\eta} + \mathcal{O}(\eta^{-3/2})\\
\end{split}
\end{equation}
This formula increases in accuracy with $\eta$, but decreases in accuracy with $n$. Since we are usually interested in the lowest energy bands $a_0$ and $a_1$, the decrease in accuracy with $n$ need not concern us.

We can use the asymptotic expansion to calculate the bandgaps $\delta_n = a_{n+1} - a_n$.
\begin{equation}
\begin{split}
\delta_n = &4\sqrt{\eta} - 1 - n\\ &- \left[\frac{3}{32} + \frac{3}{32}(2n+1) + \frac{3n^2 + 3n +1}{16}\right]\frac{1}{\sqrt{\eta}}\\
&- \left[\frac{3}{256} + \frac{2n+1}{16} + \frac{5}{128}(3n^2+3n+1)\right. \\ &+ \left. \frac{5}{256}(4n^3+5n^2+4n+1) \right]\frac{1}{\eta} + \mathcal{O}(\eta^{-3/2}).
\end{split}
\end{equation}
For the special case of the gap above the ground state this simplifies to
\begin{equation} \label{eq:gap}
\delta_0 = 4\sqrt{\eta} - 1 - \frac{1}{4\sqrt{\eta}} - \frac{17}{128\eta} + \mathcal{O}(\eta^{-3/2})
\end{equation}
which is expressed in terms of physical parameters as
\begin{equation}
\frac{\delta_0}{E_C} \approx 4\sqrt{E_J/2E_C} - 1 - \frac{1}{4\sqrt{2E_J/E_C}} - \frac{17E_C}{256E_J}.
\end{equation}
If $\eta$ is large enough that all but the $\eta^{1/2}$ terms may be neglected (physically, $E_J \gg E_C$), then this corresponds with Likharev and Zorin's result \cite{Likharev1985} based on presenting the Mathieu functions in Wannier form in the tight-binding limit, where they determine that the energy levels are just those of a harmonic oscillator
\begin{equation}
\delta_n = \hbar\omega_p
\end{equation}
where $\omega_p$ is the plasma frequency of the Josephson junction $\omega_p = \sqrt{8E_JE_C}/\hbar$.

Some quantities of physical interest are the matrix elements $A_{nm}$ of the form
\begin{equation} \label{eq:MatrixElements}
A_{nm} = \int_{-\infty}^{\infty} \textrm{d}z \psi_n^\dagger(z) A \psi_m(z)
\end{equation}
for some operator $A$.
In particular, we will be concerned with $z_{nm}$, $z_{nm}^2$, $\cos(z)_{nm}$ and $\sin(z)_{nm}$.
These can be computed analytically using the harmonic oscillator wavefunctions given in Eq.~\ref{eq:HO}.
Beginning with $z_{nm}$ we find
\begin{equation} \label{eq:znm}
\begin{split}
z_{nm} =& \int \textrm{d}z \psi_n(z)^* z \psi_m(z)\\ \approx& \frac{c_1^* c_0}{(2\eta)^{1/4}}\int \textrm{d}x  H_n^*(x) H_m(x) x e^{-x^2}.
\end{split}
\end{equation}
The matrix elements for $x = (2\eta)^{1/4}z$ are equivalent to the matrix elements of the position operator for a 1-D harmonic oscillator - an elementary calculation.
Expressing the position operator in terms of creation and annihilation operators, we find
\begin{equation}
\begin{split}
z_{nm} &= \bra{n}z\ket{m} = \eta^{-1/4}\bra{n}(a + a^\dagger)\ket{m}\\ &= \eta^{-1/4}(\sqrt{n+1}\delta_{n+1,m} + \sqrt{n}\delta_{n-1,m})
\end{split}
\end{equation}
a result which can be found in the appendix of \cite{Likharev1985}.

By the same method, we can compute the matrix element $z_{nm}^2$
\begin{equation} \label{eq:square}
\begin{split}
z_{nm}^2 =& \bra{n}z^2\ket{m} = \eta^{-1/2}\bra{n}(a + a^\dagger)^2 \ket{m}\\
=& \eta^{-1/2}\left[\sqrt{(n+1)(n+2)}\delta_{n,m-2} + \sqrt{n(n-1)}\delta_{n,m+2} \right. \\ &+ \left. 2\left(n + \frac{1}{2}\right)\delta_{n,m} \right].
\end{split}
\end{equation}

The remaining matrix elements, $\cos(z)_{nm}$ and $\sin(z)_{nm}$, cannot be so neatly expressed in terms of ladder operators (rather, each involves an infinite sum of ladder operators).
However, we can conclude that if the difference between states $|n - m|$ is odd, then $\cos(z)_{nm}$ will be zero, because $\cos(z)$ contains only even powers of the ladder operators $a$, $a^\dagger$. Similarly, if $|n-m|$ is even, then $\sin(z)_{nm}$ will be zero.
To obtain quantitative results, we can evaluate the matrix elements numerically, as is discussed in Section~\ref{sec:num}.

\section{Floquet theory and the characteristic exponent}
In physical applications, often the characteristic value is a desired output of the theory (for example in the physics of Josephson junctions it corresponds to an energy eigenvalue).
We have seen that at a given value of $\eta$ there exist continuous bands of characteristic values which give stable solutions to Mathieu's equation.
Therefore, an addition parameter is required to uniquely determine the characteristic value for a particular $\eta$.
To this end we turn to Floquet theory, where we will see that the Floquet characteristic exponent will provide the additional parameter we need.

Mathieu's equation contains periodic coefficients, so that there will exist Floquet solutions of the form
\begin{equation}
\psi(z + \tau;a,\eta) = e^{i\nu z}u_\nu(z;a,\eta).
\end{equation}
We call $u_\nu$ a Floquet solution with characteristic exponent $\nu$. These solutions are stable only if $\nu$ is real. The corresponding eigenvalues $a$ which lead to real $\nu$ form bands. This is directly analogous to the situation in solid state physics,in which allowed energy levels are precisely those which correspond to a real value for the quasi-momentum $k$ appearing in the Bloch wavefunctions for electrons in a periodic potential. Drawing out this analogy, the physical variable in Josephson junctions which corresponds to $\nu$ is called the quasi-charge, $q$.

We can explicitly include the characteristic exponent in Mathieu's equation by noting that
\begin{equation}
\frac{\partial^2}{\partial z^2}\left[e^{i\nu z}u_\nu(z)\right] = \left(\frac{\partial}{\partial z} + i\nu \right)^2 e^{i\nu z}u_\nu(z).
\end{equation}
This allows us to rewrite Mathieu's equation as
\begin{equation} \label{eq:MathieuWithNu}
\left(\frac{\partial}{\partial z} + i\nu\right)^2 \psi + \left(a - 2\eta\cos(2z)\right)\psi = 0.
\end{equation}
Stable solutions correspond to real $\nu$.

The following discussion closely follows \cite{Str/ang2005}.
The Floquet solutions are periodic, so they can be expanded as a Fourier series
\begin{equation} \label{eq:Floquet}
\psi(z,a) = e^{i\nu z}\sum_{\kappa  \in \mathbb{Z} } c_{2\kappa}(\nu;a,\eta)e^{2i\kappa z}.
\end{equation}
If we insert this expansion into Eq.~\ref{eq:Mathieu} we obtain a three-term recursion formula for the coefficients
\begin{equation}
\left((2\kappa - \nu)^2-a \right) c_{2\kappa} + \eta(c_{2(\kappa+1)} + c_{2(\kappa-1)}) = 0, \quad \forall\kappa \in \mathbb{Z}
\end{equation}
which we will re-write as
\begin{equation}
c_{2\kappa} + \frac{\eta\left(c_{2(\kappa + 1)} + c_{2(\kappa - 1)}\right)}{(2\kappa - \nu)^2 - a} = 0, \quad \forall\kappa \in \mathbb{Z}.
\end{equation}
If we choose a truncated upper limit $n \in \mathbb{Z}$ then the recursion relation can be written as a matrix equation
\begin{equation}
A_n(\nu;a,\eta)\vec{c}_n = 0
\end{equation}
where
\begin{widetext}
\begin{equation}
A_n(\nu;a,\eta) = \begin{pmatrix}
1 & \xi_{2n} & 0 & \dots &&&& \dots & 0 \\
\xi_{2n-2} & 1 & \xi_{2n-2} & \ddots &&&&&\\
0 & \ddots & \ddots & \ddots &&&&&\\
\vdots && \xi_2 & 1 & \xi_2  &&&&\\
&&&\xi_0 & 1 & \xi_0 &&&\\
&&&& \xi_{-2} & 1& \xi_{-2} &&\\
&&&&& \ddots & \ddots & \ddots &\\
\vdots &&&&&&& 1 & \xi_{-2n+2}\\
0 &&&&&&& \xi_{-2n} & 1
\end{pmatrix}
\end{equation}
\end{widetext}
and
\begin{equation}
\xi_{2\kappa} = \frac{\eta}{(2\kappa - \nu)^2 -a}
\end{equation}
In finite dimensions, finding non-trivial solutions to $A_n(\nu;a,\eta)\vec{c}_n = 0$ is equivalent to demanding that $\det{A_n(\nu;a,\eta)} = 0$. For convenience, we will introduce the notation $\Delta(a,\nu) = \det{A_n(\nu;a,\eta)}$. We shall now examine some properties of $\Delta(a,\nu)$.

Note that $(2\kappa - (\nu + 2)) = (2(\kappa - 1)-\nu)$, so that $\xi_{2\kappa}(\nu+1) = \xi_{2(\kappa+1)}(\nu)$. Since $\kappa$ takes all values from $-\infty$ to $+\infty$, this gives us $\Delta(a,\nu) = \Delta(a,\nu+1)$, showing us that $\Delta(a,\nu)$ is a periodic function in $\nu$ with period 1. This means we can restrict our analysis to the strip $0\leq\nu\leq1$.

The functions $\xi_{2\kappa}$ have simple poles at values of $a$ and $\nu$ which satisfy $2\kappa - \nu = \pm\sqrt{a}$. Apart from these poles, $\Delta(a,\nu)$ is analytic. We can avoid these poles by constructing a function
\begin{equation}
D(a,\nu) = \frac{1}{\cos(\pi\nu) - \cos(\pi\sqrt{a})}
\end{equation}
which has poles at the same values of $a$ and $\nu$ as $\Delta(a,\nu)$. Now if we choose an appropriate function $C(\nu)$, the function
\begin{equation}
\Theta(a,\nu) = \Delta(a,\nu) - C(\nu)D(a,\nu)
\end{equation}
has no singularities. Because we can confine ourselves to the strip $0\leq\nu\leq 1$, we only have one pole to worry about in $\Delta(a,\nu)$ and $D(a,\nu)$, so finding $C(\nu)$ is just the problem of calculating the constant $C$ corresponding to the quotient between residuals of $\Delta(a,\nu)$ and $D(a,\nu)$.

If we chose $C$ as to correctly eliminate singularities, then $\Theta(a,\nu)$ is an analytic function in the entire complex plane with no poles. By Liouville's theorem, it must therefore be a constant. 

In the limit $\nu\rightarrow+i\infty$, all of the $\xi_{2\kappa}$ functions go to zero and the matrix $A(\nu;a,q)$ is reduced to the identity matrix. In this limit, therefore, $\Delta(a,\nu) = 0$. Furthermore, $\cos(\pi\nu)\rightarrow\infty$, and $D(a,\nu)\rightarrow0$. Hence $\Theta(a,\nu) = 1$, and
\begin{equation}
C = \frac{\Delta(a,\nu) - 1}{D(a,\nu)}.
\end{equation}
Because $C$ is a constant, this relation must always hold for any $a$ and $\nu$.
In the case $\nu = 0$, $D(a,\nu) = 1/(1-\cos(\pi\sqrt{a}))$. This gives us
\begin{equation}
\begin{split}
C =& \left(\Delta(a,0) - 1\right)\left(1 - \cos(\pi\sqrt{a})\right)\\ =& 2(\Delta(a,0) - 1)\sin^2\left(\frac{\pi\sqrt{a}}{2}\right)
\end{split}
\end{equation}
For more general values of $a$ and $\nu$, we are seeking cases where $\Delta(a,\nu) = 0$. Hence,
\begin{equation}
C = \cos(\pi\sqrt{a})-\cos(\pi\nu).
\end{equation}
Equating these two expressions, we obtain the Whittaker-Hill formula \cite{Whittaker1915}
\begin{equation} \label{eq:WittHill}
\sin^2\left(\frac{\pi\nu}{2}\right) = \Delta(a,0)\sin^2\left(\frac{\pi\sqrt{a}}{2}\right)
\end{equation}
which gives us a relation between the characteristic exponents of the Floquet solutions to Mathieu's equation and their corresponding eigenvalues.
Some of the literature uses the equivalent formula
\begin{equation}
\cosh\left(i\nu\pi\right) = 1 - 2\Delta(a,0)\sin^2\left(\frac{\pi\sqrt{a}}{2}\right).
\end{equation}

\subsection{Evaluating $\Delta_0$}
Since $\Delta_0$ is the determinant of an infinite matrix, exact evaluation is not possible except for in the limit $\eta \rightarrow 0$, in which case $\Delta_0 = 1$.
To calculate $\Delta_0$ numerically, we must truncate it at some finite cut-off $n$.
We denote the determinant of the truncated matrix as $\Delta_n$.
To calculate the determinant of the full infinite-dimensional matrix $\mathcal{A}(0;a,\eta)$ we will need to limit $n \rightarrow \infty$, however for small $\eta$ we will be able to obtain a good approximation at readily achievable finite values of $n$.

A formula for obtaining $\Delta_n$ in terms of determinants at smaller truncations in given in \cite{Str/ang2005} as
\begin{equation} \label{eq:recursion}
\Delta_n = (1 - \alpha_n)\Delta_{n-1} - \alpha_n(1 - \alpha_n)\Delta_{n-2} + \alpha_n\alpha_{n-1}^2\Delta_{n-3},
\end{equation}
where $\alpha_n = \xi_{2n}\xi_{2n-2}$.
We take as a starting point $\Delta_{<0} = 0$, $\Delta_0 = 1$ and $\Delta_1 = (1 - 2\alpha_1)$ (which, as the determinant of a $3\times 3$ matrix, can be easily calculated by hand). 

Note that $\Delta_n$ is the determinant of a $2n+1 \times 2n+1$ matrix.
For $n=1$ we have
\begin{equation}
\Delta_1(a,\eta) = 1 - 2\xi_0\xi_2 = 1 - \frac{2\eta^2}{a(a-4)}
\end{equation}
and at $n=2$
\begin{align}
\begin{split}
\Delta_2(a,\eta) & = 1 - 2\xi_0\xi_2 -  2\xi_2\xi_4 +2\xi_0\xi_2^2\xi_4 + \xi_2^2\xi_4^2\\ &= 1 - \frac{4(a-8)\eta^2}{a(a^2 - 20a + 64)} + \frac{(3a - 32)\eta^4}{a(a^2 - 20a + 64)^2}.
\end{split}
\end{align}

The $\mathcal{O}(\eta^2)$ changes when we move from $\Delta_1$ to $\Delta_2$. 
This iterative approach is therefore not the same as expanding in powers of $\eta$, even though a new power of $\eta$ will appear at each increment of $n$.
For small $\eta$, it is reasonable to neglect higher powers of $\eta$.
If we continue to apply Eq.~\ref{eq:recursion}, but discard all terms of order $\eta^4$ or higher, we obtain
\begin{equation}
\Delta_N = 1 - 2\sum_{k=1}^{N}\alpha_k = 1 - 2\sum_{k=0}^{N}\frac{\eta^2}{[(2k)^2 - a][(2k-2)^2 - a]}.
\end{equation}
It should be noted that since $\Delta_n$ is evaluated at $\nu=0$, the value of $a$ entering the equation is always the minimum of the band.
Therefore $a$ is strictly negative unless $\eta = 0$, so the above sum does not contain any poles.

\section{Comparison to numerical solutions} \label{sec:num}
Despite the lack of exact analytic results, numerically solving Mathieu's equation is quite straight-forward.
In the application of Mathieu's equation to Josephson junction arrays we make use of the fact that phase and charge are canonical conjugate variables, and re-write Schr\"{o}dinger's (Mathieu's) equation (Eq.~\ref{eq:MathieuWithNu}) in the charge basis:
\begin{eqnarray}
\sum_n &\left[-4E_C(\hat{n} - q)^2 - \frac{E_J}{2}\left(\ket{n+1}\bra{n} + \ket{n-1}\bra{n}\right)\right]\ket{\psi_m}\nonumber \\ &= E_m \ket{\psi_m},
\end{eqnarray}
where $\hat{n}$ is the Cooper pair number operator, $m$ labels the different energy levels and $q$ is the quasicharge (corresponding to $\nu$ in the generic Mathieu equation notation).
While theoretically the sum over $n$ should run over $n \in (-\infty,\infty)$, for practical purposes we must truncate at some finite value $N$.
With the correct translation between Mathieu function and Josephson junction notation, the discretized Mathieu's equation is reduced to the problem of finding the eigenvalues and eigenvectors of the $(2N+1)\times(2N+1)$ tridiagonal matrix

\begin{widetext}
\begin{equation} \label{eq:matrix}
\begin{pmatrix}
(\nu + N)^2 & -\eta & & & &\\
-\eta & (\nu + N - 1)^2 & -\eta & & &\\
& \ddots & \ddots  & \ddots & &\\
& & -\eta & (\nu)^2 & -\eta &\\
& & & \ddots & \ddots & \ddots\\
& & & & -\eta & (\nu - N + 1)^2 & -\eta\\
& & & & & -\eta & (\nu - N)^2
\end{pmatrix}\\
\end{equation}
\end{widetext}

We solve this eigenvalue equation for each value of $\nu$ separately.
The resulting eigenvectors are discrete Mathieu's functions, and the resulting eigenvalues are $a_m(\nu)$.
Other quantities of interest can be obtained from these, for example the bandwidth $b_1 - a_0$ can be obtained numerically as $\max[a(\nu)] - \min[a(\nu)]$ (where $a(\nu)$ is the continuous band of characteristic values, of which $b_1$ and $a_0$ are the maximum and minimum values respectively).

Since the numerical results can be calculated to arbitrary precision, we use these to test the validity of the analytic approximations introduced above.

\begin{figure}[b!]
	\centering
	\includegraphics[width=1\linewidth]{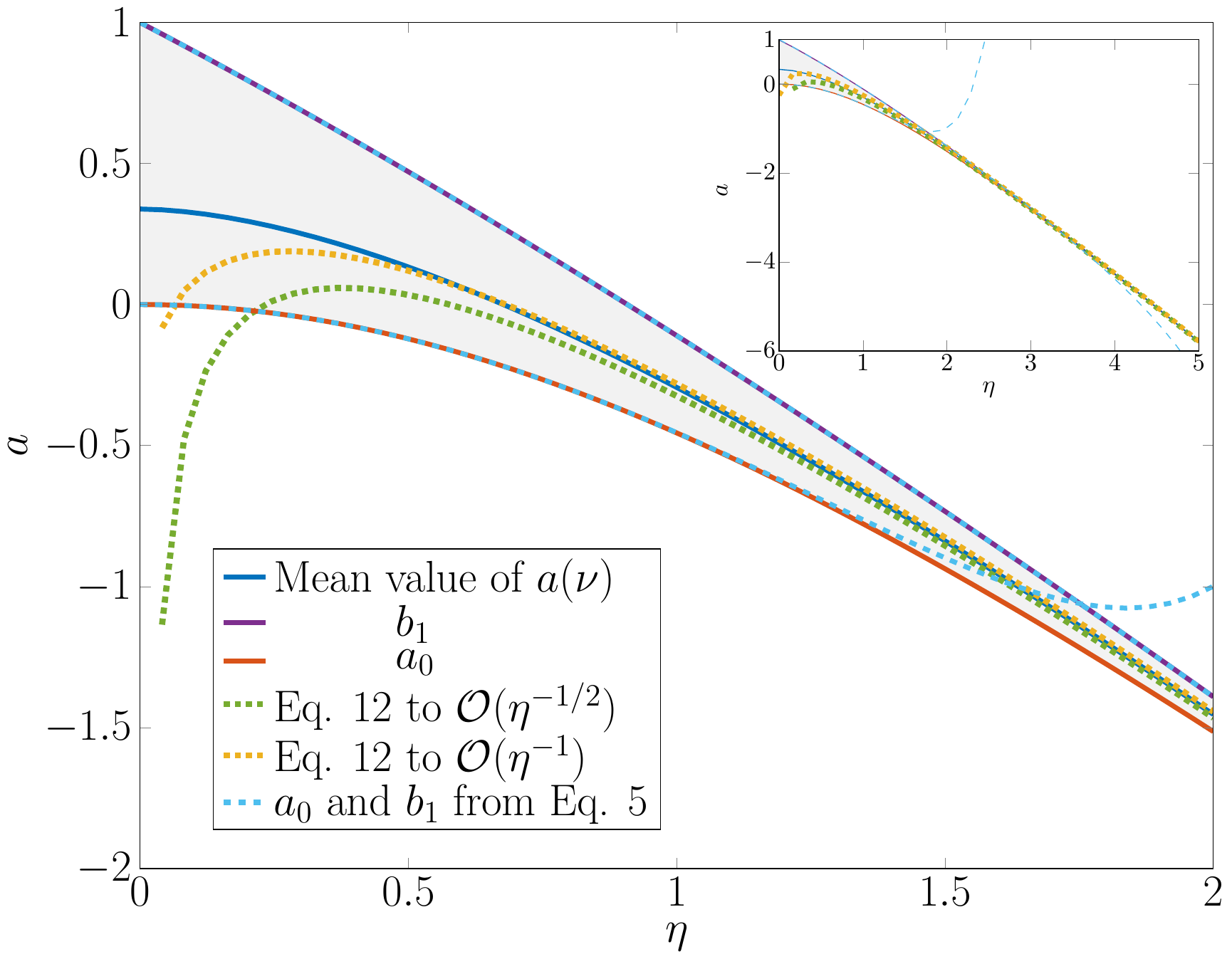}
	\caption{\label{fig:Eigenvalues} Comparison between various methods for computing characteristic values at small $\eta$. Solid lines correspond to numerical results, while dashed lines show various analytic approximations. It can be seen that asymptotic expansions expected only to be valid for $\eta \gg 1$ already well approximate the mean of $a(\nu)$ at $\eta \sim 1$. Furthermore, for $\eta \lesssim 1$, the small $\eta$ expansions of $a_0$ and $b_1$ are indistinguishable from numerical results.}
\end{figure}

Fig.~\ref{fig:Eigenvalues} displays the characteristic values $a_0$ and $b_1$, as well as the mean value of the lowest band of $a(\nu)$ for small $\eta$, giving a direct comparison between the asymptotic expansions given in Eq.~\ref{eq:McLachlan} (valid for small $\eta$), Eq.~\ref{eq:F+P} (valid for large $\eta$) and numerical calculations (valid to arbitrary precision across all values of $\eta$). We see that Eq.~\ref{eq:McLachlan} is extremely accurate up to $\eta = 1$, at which point the approximate value of $a_0$ begins to diverge significantly from the numerically calculated value.

In the inset of Fig.~\ref{fig:Eigenvalues} we plot the same characteristic values to higher values of $\eta$, and observe the gap between $a_0$ and $b_1$ shrinking exponentially. Approximating these two quantities as a single value given by Eq.~\ref{eq:F+P} becomes more and more accurate at larger values of $\eta$.

\begin{figure}
	\centering
	\includegraphics[width=1\linewidth]{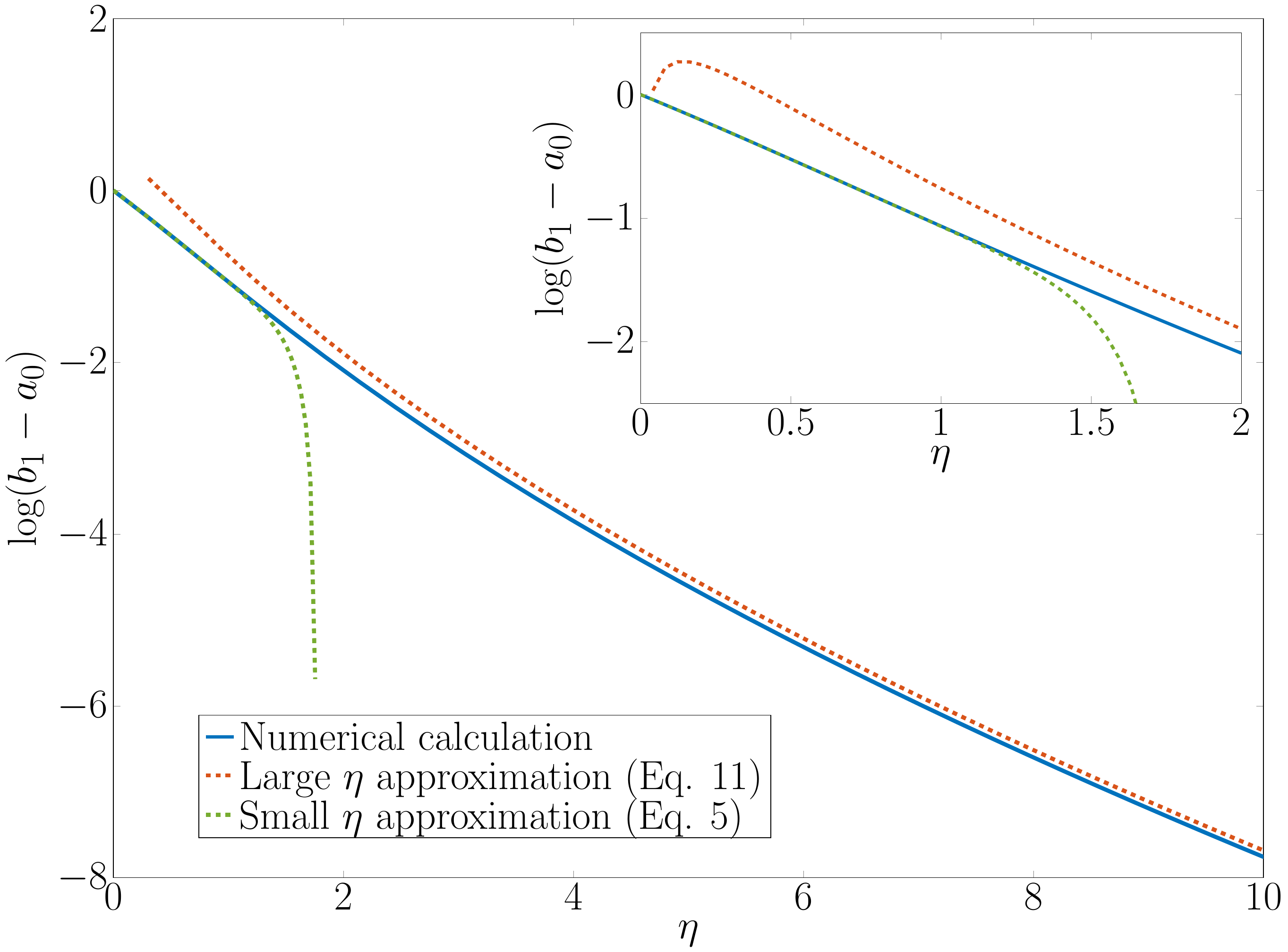}
	\caption{\label{fig:Bandwidth} At small $\eta$ the bandwidth $b_1 - a_0$ can be reliably calculated from power law expansions of the characteristic values given in section 2.151 of \cite{McLachlan1947}. The tight-binding approximation of the bandwidth remains inadequate until $\eta \gg 1$. }
\end{figure}

The exponential suppression of the bandwidth is demonstrated in Fig.~\ref{fig:Bandwidth}, where the natural logarithm of the numerically calculated value $b_1 - a_0$ is compared to the small $\eta$ approximation calculated from Eq.~\ref{eq:McLachlan} and the tight-binding approximation Eq.~\ref{eq:TBbandwidth}.
The bandwidth is shown to higher values of $\eta$ in the inset of Fig.~\ref{fig:Bandwidth}, where it can be seen that Eq.~\ref{eq:TBbandwidth} becomes a good approximation at large $\eta$.

\begin{figure}[tb!]
	\centering
	\includegraphics[width=1\linewidth]{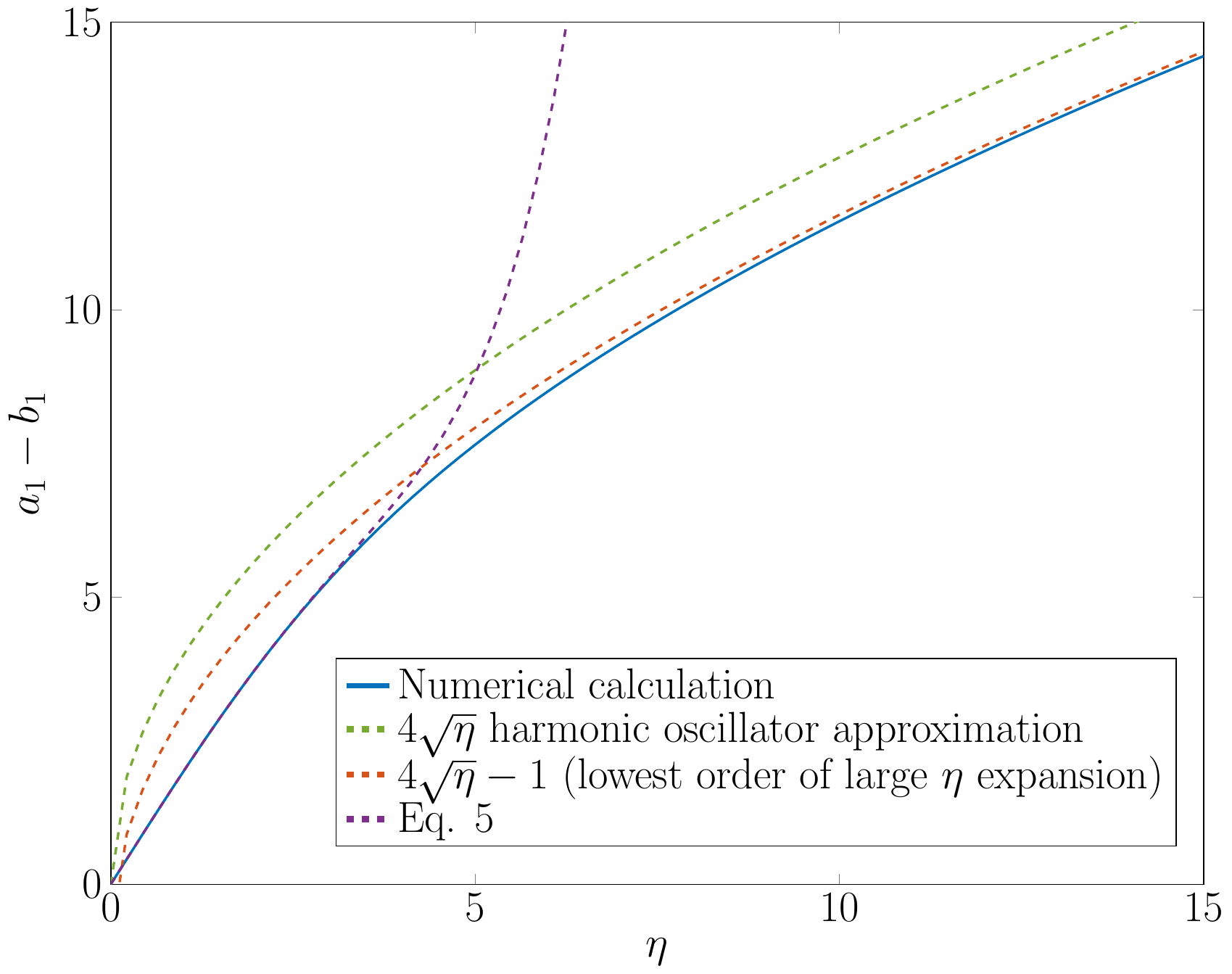}
	\caption{\label{fig:Bandgap} The gap between stable solutions of Mathieu's equation is equivalent to the gap between allowed energy levels in a Josephson junction. Here we see that the power law expansions given in section 2.151 of \cite{McLachlan1947} fit the numerically calculated value very well when $\eta \lesssim 1$, but rapidly diverge at higher values. The harmonic oscillator value $4\sqrt{\eta}$ often quoted in the physics literature is only a good fit for values of $\eta$ much larger than those presented here. A first order correction derived from the asymptotic expansions of \cite{Frenkel2001} produces a much better fit for intermediate values of $\eta$. }
\end{figure}

Fig.~\ref{fig:Bandgap} shows the width of the gap between characteristic values which correspond to stable solutions of Mathieu's equation, physically corresponding to the band gap between the first two energy bands of a Josephson junction.
The approximation $a_1 - b_1 = 4\sqrt{\eta}$ is common in the literature on Josephson junctions.
This is simply Eq.~\ref{eq:gap} to lowest order, $\mathcal{O}(\eta^{1/2})$.
Going to the next highest order ($0^\textrm{th}$ order) is trivial, and yields a major improvement to the accuracy of the approximation.
Eq.~\ref{eq:McLachlan} and the surrounding text can give us a small $\eta$ approximation for $a_1 - b_1$ which is very accurate for $\eta \lesssim 1$.

\begin{figure}[b!]
	\centering
	\includegraphics[width=1\columnwidth]{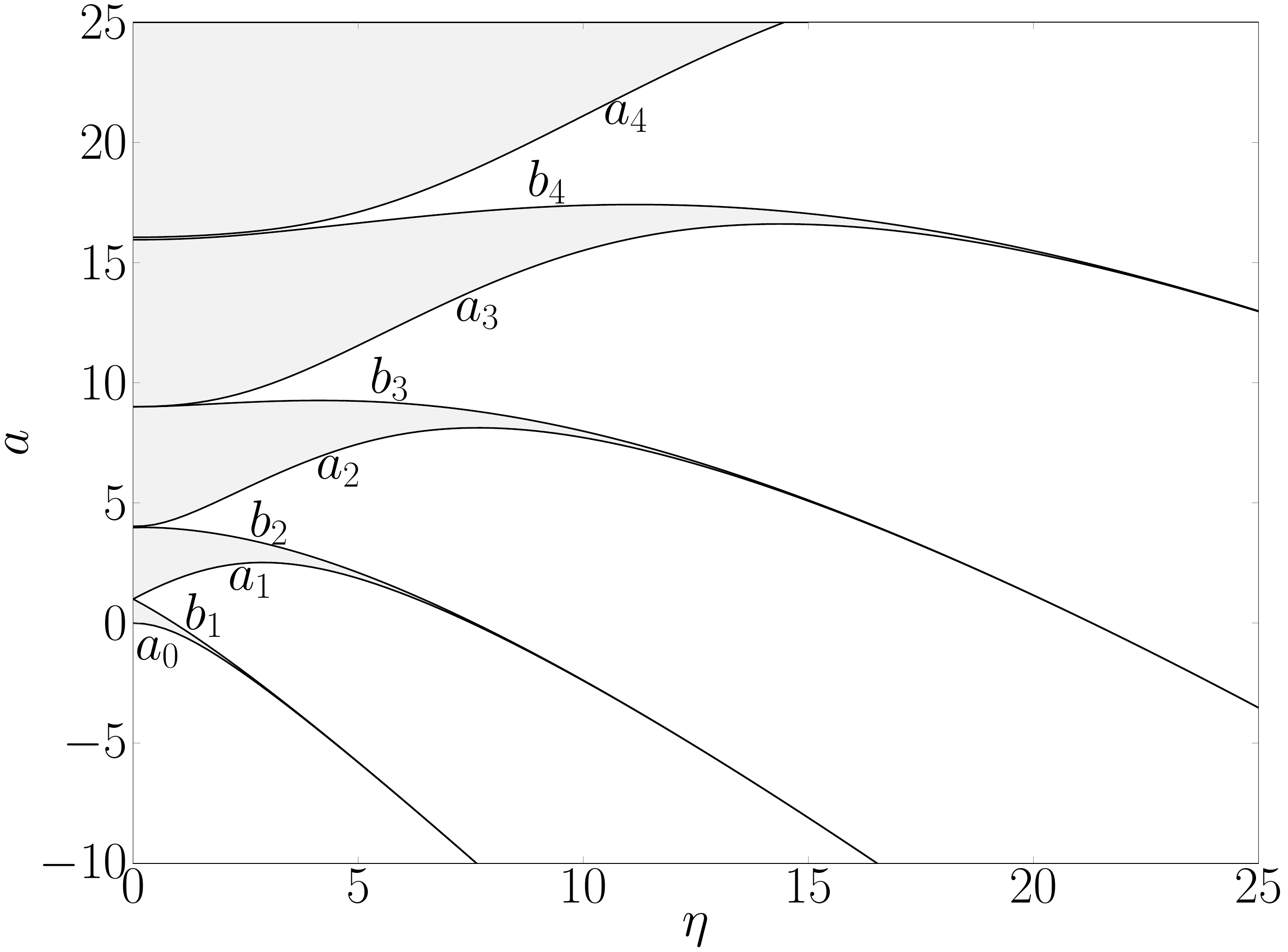}
	\caption{\label{fig:Stability} Shaded regions between the curves $a_n$ and $b_{n+1}$ correspond to stable solutions of Mathieu's equation. The stability diagram of Mathieu's equation is equivalent to the band diagram of a Josephson junction. Here we can see visually the exponential suppression of bandwidths with increased coupling $\eta$. }
\end{figure}

Finally, Fig.~\ref{fig:Stability} shows which values of $a$ and $\eta$ correspond to stable solutions of Mathieu's function. 
This corresponds to the band structure of a Josephson junction. 
The characteristic values, bandwidths and bandgaps plotted in previous figures can be seen together here.

If we wish to make use to the matrix representation in Eq.~\ref{eq:matrix} to calculate the matrix elements $z_{nm}$ numerically, we must change the basis of our wavefunctions
\begin{equation}
\psi_n(z) = \frac{1}{\sqrt{2\pi}}\sum_k \psi_n^k e^{ikz}
\end{equation}
where the superscript $k$ is an index labelling the basis vector, not a power.
inserting this into our definition of $z_{nm}$ in Eq.~\ref{eq:znm} we obtain

\begin{widetext}
\begin{equation}
\begin{split}
z_{nm} &= \frac{1}{2\pi} \int_{-\pi}^{\pi} dz \sum_{k,q} (\psi_n^k)^\dagger e^{-ikz} z e^{iqz}\psi_m^q
= \frac{1}{2\pi} \sum_{k,q} (\psi_n^k)^\dagger \psi_m^q \int_{-\pi}^{\pi} dz ze^{i(q-k)z}\\
&= \frac{1}{2\pi} \sum_{k,q} (\psi_n^k)^\dagger \psi_m^q \frac{2i\left(\sin(\pi(k-q)) -\pi(k-q)\cos(\pi(k-q)) \right)}{(k-q)^2}
\end{split}
\end{equation}
\end{widetext}

except for in the case $k=q$, in which case the integral evaluates to $0$.
Since $k-q$ can only take integer values, $\sin(\pi(k-q)) = 0$ and $\cos(\pi(k-q)) = (-1)^{k-q}$.
Introducing variable $p = k-q$, the above expression simplifies to
\begin{equation} \label{eq:z_nm_num}
z_{nm} = \sum_{k,q} (\psi_n^k)^\dagger \psi_m^{k-p} \frac{i(-1)^p}{p}.
\end{equation}
A similar calculation for $z_{nm}^2$ gives
\begin{equation} \label{eq:square_num}
z_{nm}^2 = 2 \sum_{k,q} (\psi_n^k)^\dagger \psi_m^q \frac{(-1)^p}{p^2},
\end{equation}

\begin{figure}
	\centering
	\includegraphics[width=1\linewidth]{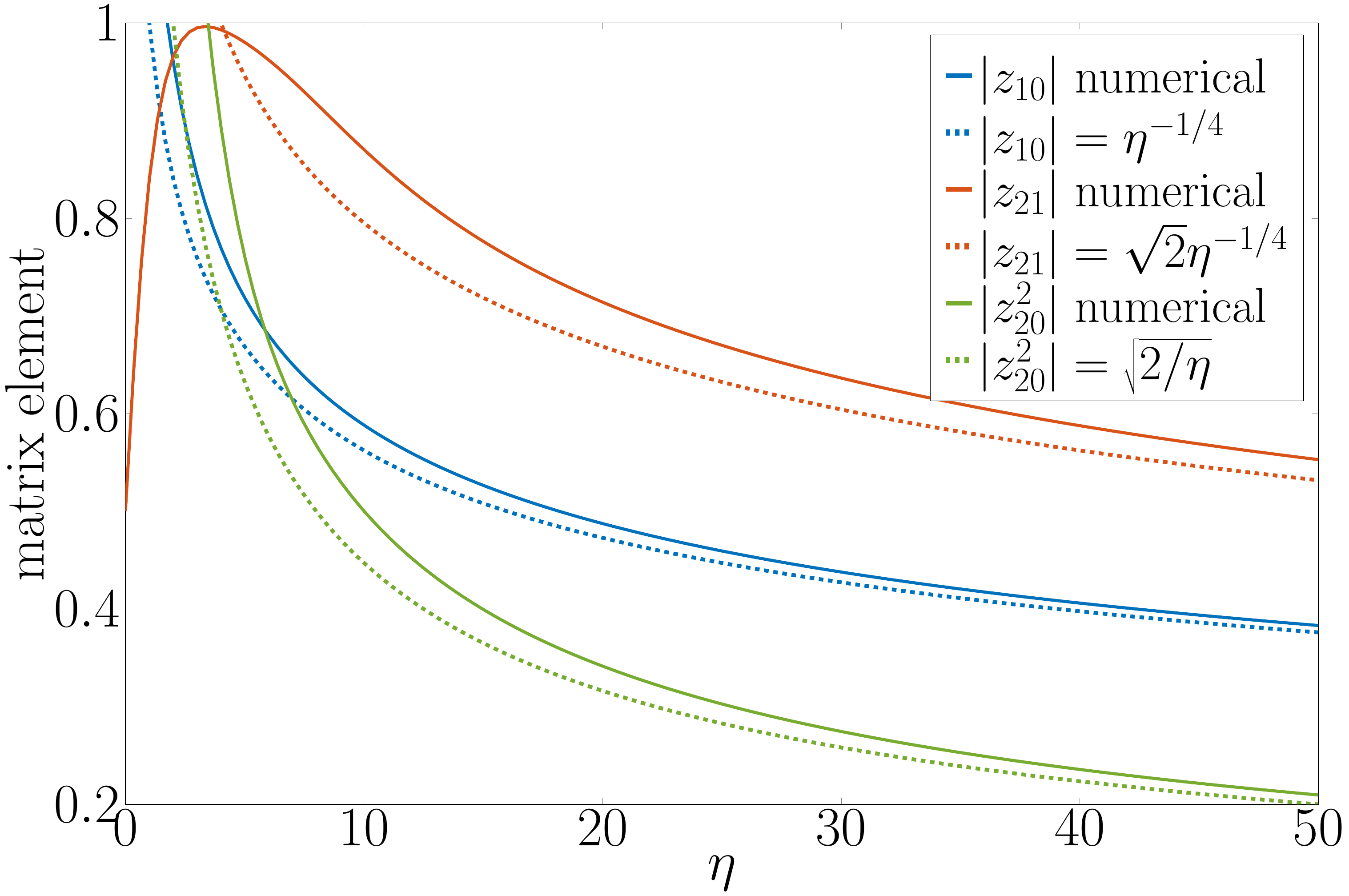}
	\caption{\label{fig:z_nm} Matrix elements $|z_{nm}|$ ans $|z^2_{nm}|$ calculated using the numerical methods of Eqs.~\ref{eq:z_nm_num} and \ref{eq:square_num} (solid lines) and the analytic approximations Eqs.~\ref{eq:znm} and \ref{eq:square} (dashed lines).}
\end{figure}

With numerically obtained vectors $\psi_n^k$, we can calculate the above expressions and compare it to the expressions we obtained analytically by approximating the Mathieu functions as wavefunctions of the harmonic oscillator, as is done in Fig.~\ref{fig:z_nm}.

Other matrix elements of interest are $\cos(z)_{nm}$ and $\sin(z)_{nm}$.
These can be easily computed numerically using the representations
\begin{equation}
\begin{split}
\cos(z)_{nm} = \frac{1}{2} \sum_p \bra{n}\left( \ket{p+1}\bra{p-1} + \ket{p-1}\bra{p+1} \right)\ket{m}\\
\sin(z)_{nm} = \frac{i}{2} \sum_p \bra{n}\left( \ket{p+1}\bra{p-1} - \ket{p-1}\bra{p+1} \right)\ket{m}.
\end{split}
\end{equation}

\section{Effective voltage}
\begin{figure}
	\centering
	\includegraphics[width=1\linewidth]{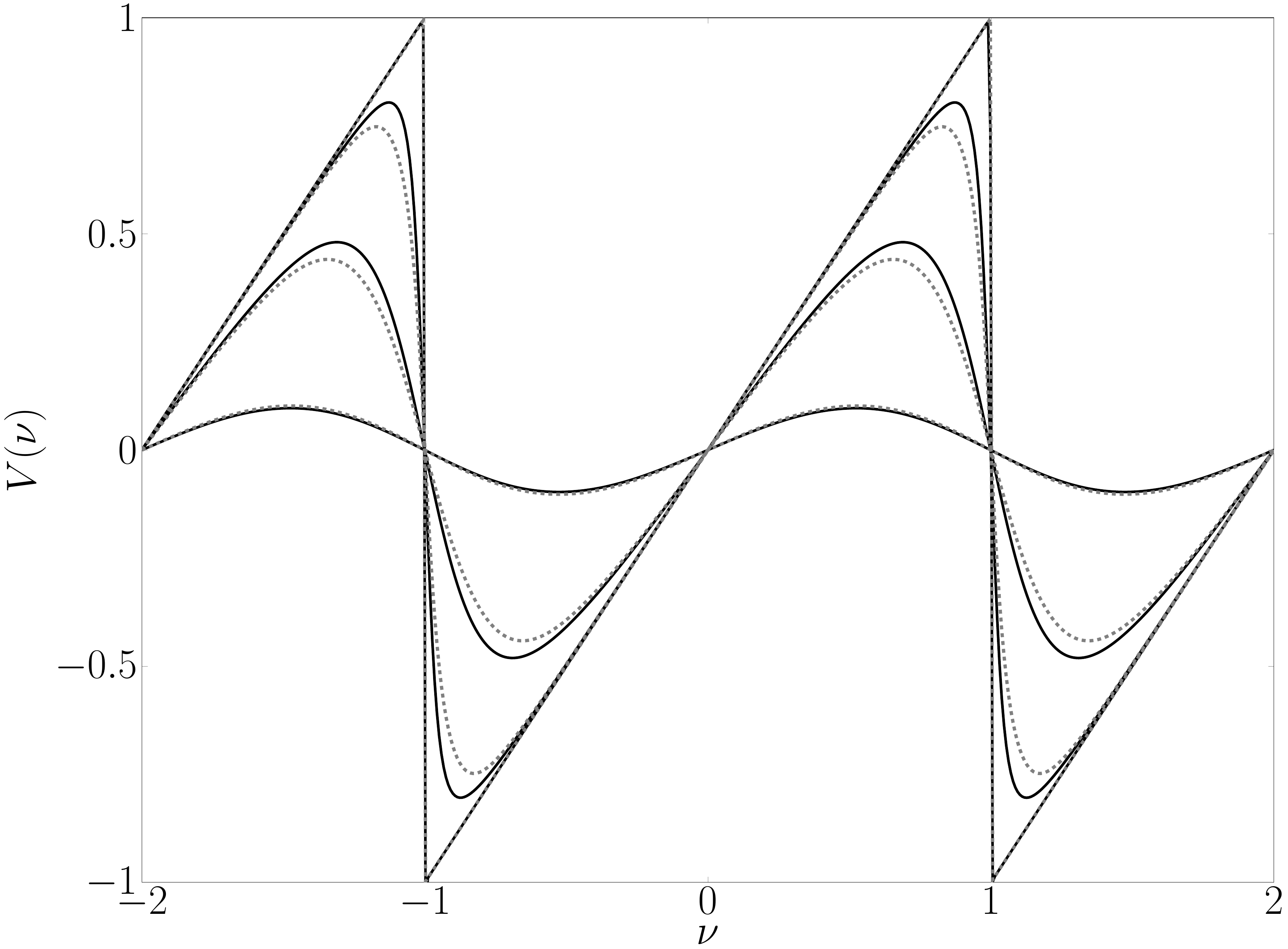}
	\caption{\label{fig:Dima} Eq.~\ref{eq:Dima} for the effective voltage $V(\nu)$ (dashed lines) compared with numerical calculations (solid lines) for $\eta = 0, 0.1, 0.5$ and $2$. It can be seen that the approximation matches the numerical results extremely well, especially for large $\eta$ and for $\eta = 0$. Solid lines correspond to the numerically calculated values, while the dashed lines are calculated using Eq.~\ref{eq:Dima}.}
\end{figure}

When $E_C \sim E_J$, it is convenient to describe the Josephson junction not in terms of discrete charges $n$ or in terms of the Josephson phase $\phi$, but rather in terms of the quasicharge $q$ \cite{Likharev1985} (equivalent to the characteristic exponent $\nu$ of Mathieu's equation).
In this case, the effective voltage across a junction is $dE_0/dq$, or, in the Mathieu equation notation used above, $da/d\nu$.
Our asymptotic formulae above approximate the bands of $a$ as infinitely thin in $\nu$, and therefore do not include explicit $\nu$ dependence.
Instead, a semi-analytic approach has previously been employed (as presented in the thesis of Adem Erg\"{u}l \cite{Ergul2013a}), where the function form of $V(\nu)$ has been obtain from the relation Eq.~\ref{eq:WittHill} and constants have been chosen so as to reproduce the correct limits as $\eta \rightarrow 0$ and $\eta \rightarrow \infty$.
This approach gives us
\begin{equation} \label{eq:Dima}
V(\nu) = \frac{4}{\pi}\arcsin\left(\frac{\sin(\nu)}{\sqrt{f+2}\sqrt{f+1+\cos(\nu)}}\right),
\end{equation}
where $f \cong 1.2\eta^2$ is a fitting parameter chosen to give the correct results in the limits $\eta \rightarrow 0$ and $\eta \rightarrow \infty$.
(The use of such a parameter is made necessary due to the difficulty in analytically evaluating the infinite determinant $\Delta_0$.)
This functional form is a very good approximation across all values of $\eta$, matching numerical calculations very closely, as can be seen in Fig.~\ref{fig:Dima}.

\section{Conclusion}
Mathieu's equation appears in many problems within theoretical physics.
In most situations, it is convenient to simply solve the equation numerically.
In some cases, however, an analytic approximation may be desired.
We have gathered here several analytic approximations for various quantities relating to Mathieu's equation and compared them to numerical results (which may, in princple, be evaluated to arbitrary accuracy).

One results of particular interest is that of the gap between stable solutions of Mathieu's equation - physically corresponding to a bandgap.
In much of the physics literature the characteristic values of Mathieu's equations are approximated as the eigenvalues of a harmonic oscillator (see, for example \cite{Likharev1985}). In this paper we have seen that the harmonic oscillator approximation corresponds to a first order approximation with respect to the low $\eta$ asymptotic expansions of Frenkel and Portugal \cite{Frenkel2001}.
Extending the approximation to second order is trivial - it merely involves an additive constant - but already yields a large improvement to the approximation, as can be seen in Fig.~\ref{fig:Bandgap}.

\section{Acknowledgements}
This work was supported in part by the Australian Research Council under the Discovery and Centre of Excellence funding schemes (project numbers DP140100375 and CE170100039). Computational resources were provided by the NCI National Facility systems at the Australian National University through the National Computational Merit Allocation Scheme supported by the Australian Government.

\bibliographystyle{unsrt}
\bibliography{Bib}
\end{document}